\documentclass[aps,prl,preprint]{revtex4-1}
\usepackage{amssymb,amsmath,bm}
\usepackage{graphicx}
\usepackage{multirow}
\usepackage{hyperref}
\hypersetup{colorlinks=true,citecolor=blue,linkcolor=red,urlcolor=blue}
\usepackage{color}

\definecolor{green}{rgb}{0,0.5,0}

\begin{document}
\title{Unjamming of active rotators}
\author{Linda Ravazzano \textit{$^{a}$}, Maria Chiara Lionetti\textit{$^{b}$}, Maria Rita Fumagalli\textit{$^{b,c}$}, Silvia Bonfanti\textit{$^{a}$}, Roberto Guerra\textit{$^{a}$},  Oleksandr Chepizhko \textit{$^{d}$},  Caterina A. M. La Porta \textit{$^{\ast,b,c,e}$}, Stefano Zapperi\textit{$^{\ast,a,f}$}, }

\affiliation{\textit{$^{a}$Center for Complexity and Biosystems,
Department of Physics, University of Milano, via Celoria 26, 20133 Milano, Italy }}
\affiliation{\textit{$^{b}$ Center for Complexity and Biosystems,
Department of Environmental Science and Policy,  University of Milan, via Celoria 26, 20133 Milano, Italy}}
\affiliation{\textit{$^{c}$ CNR - Consiglio Nazionale delle Ricerche, Biophysics institute, via De Marini 6, Genova, Italy}}
\affiliation{\textit{$^{d}$ Institut f\"ur Theoretische Physik, Leopold-Franzens-Universit\"at Innsbruck, Technikerstrasse 
21a, A-6020 Innsbruck, Austria}}
\affiliation{\textit{$^{e}$ Innovation for Well-Being and Environment (CR-I-WE), University of Milan, Milan, Italy.}}
\affiliation{\textit{$^{f}$ CNR - Consiglio Nazionale delle Ricerche,  Istituto di Chimica della Materia Condensata e di Tecnologie per l'Energia, Via R. Cozzi 53, 20125 Milano, Italy}}
\affiliation{$^\ast$ Corresponding authors: caterina.laporta@unimi.it,stefano.zapperi@unimi.it}

\begin{abstract}
Active particle assemblies can exhibit a wide range of interesting 
dynamical phases depending on internal parameters such as density, adhesion strength
or self-propulsion. Active self-rotations are rarely studied in this context, although
they can be relevant for active matter systems, as we illustrate by analyzing the motion
of {\it Chlamydomonas reinhardtii} algae under different experimental conditions.
Inspired by this example, we simulate the dynamics of a system of interacting active disks 
endowed with active torques. At low packing fractions, adhesion causes the formation of small rotating clusters, resembling those observed when algae are stressed.
At higher densities, the model shows a jamming to unjamming transition
promoted by active torques and hindered by adhesion. Our results yield
a comprehensive picture of the dynamics of active rotators, providing useful 
guidance to interpret experimental results in cellular systems where rotations
might play a role.

\end{abstract}

\maketitle
 
\section{Introduction} 
Recent years witnessed a growing interest in the properties of active matter, where  
the system is composed by self-propelled units. The field was inspired by the observation of natural phenomena such as the movements of flocks of birds, schools of fishes or cell population, and pushed forward by new technological achievements that allowed the production of active colloids, Janus particles and artificial microswimmers. Basic concepts of statistical physics have been applied to the study of ensembles of active particles in order to investigate not only  single particle properties, but also their collective behavior. Active matter has shown a 
rich variety of emerging phenomena, due to the fact that those systems are out of thermodynamical equilibrium. Notable examples are the emergence of vortices in confined geometries \cite{bricard2015emergent}, the formation of clusters \cite{levis2014clustering} and active cristals \cite{bechinger2016active}, and phase transitions that differ from the ones typical of passive particles. For instance in active particle systems, the activity itself can induce phase separation  \cite{digregorio2018full} or phase coexistence between regions with hexatic order and regions in the liquid or gas phase \cite{cugliandolo2017phase}.

Most theoretical studies of active matter consider self-propelled particles driven by active 
traslational forces. The observation of biological active matter suggests, however, that  active rotations may also play an important and yet unexplored role. An interesting example
is provided by \textit{Chlamydomonas reinhardtii} (\textit{C. reinhardtii}), a micron-sized unicellular alga that is able to move thanks to two flagella. It has been noticed that those organisms not only self-propel to perform translational motion, but they also have the ability to self-rotate. This peculiar behavior is due to the morphology of this alga, characterized by an 'eyespot' sensible to light located near the cell equator. Rotating around its own axis the alga allows the eyespot to better scan the surrounding environment looking for light, needed to perform photosyntesis \cite{choudhary2019reentrant}. In this paper, we report observations and quantification of individual rotation and the formation of rotating clusters, in analogy to what observed for model systems of active rotating disks in 2D passive media \cite{aragones2019aggregation}. 

Inspired by this biological example, we study the dynamic behavior of a collection of interacting active rotators in two dimensions. Our model system displays some analogies with chiral active fluids which are composed by particles spinning with a defined chirality and display peculiar physical properties such an odd (or Hall) viscosity due to breaking both parity and time-reversal symmetries \cite{banerjee2017odd} and the emergence of active turbulence behavior \cite{kokot2017active}. Here we concentrate our attention on a non-equilibrium phase transition that such system of active rotators can display: the jamming-unjamming transition. 
This transition, typical of granular materials, shares some features
with the glass transition. Increasing the density, the system goes from an unjammed liquid-like phase to a jammed solid-like state, characterized by limited mobility and slow relaxation \cite{liu1998jamming}. Interestingly, this kind of transition has been observed to take place also in systems of living cells where it may play a role in biological processes, such as inside the epithelial tissues of patients affected by asthma \cite{park2015unjamming} or in the migration of cancer cells during wound healing  \cite{chepizhko2018jamming}. In this paper, we explore
the role of self-rotation on the jamming-unjamming phase transition of a system of bidimensional disks, performing molecular dynamics simulations with LAMMPS (http://lammps.sandia.gov.) \cite{plimpton1993fast}. 

\section{Materials and Methods \label{sec:methods}}
\subsection{Experiments}
\subsubsection{ \textit{C. reinhardtii} culture growth and exposure to stress conditions.}
\textit{C. reinhardtii} cells were growth in TAP medium (Invitrogen) as batch cultures until they reached $1-2 \times 10^6$ cells/ml (corresponding to mid-exponential phase of growth). The cells were cultured under continuous cool-white fluorescent lamps ($\simeq 100\mu$ mol photons/m$^2$ s) within a $110$ rpm shaking incubator, at $25^\circ$C. For palmelloid analysis, $5$ ml of cells was spun at $1100$ g/5 min/25$^\circ$ C and resuspended in $20$ ml of Tris-Acetate-Phosphate (TAP, Invitrogen cod. A1379801) medium containing $150$ mM NaCl or in fresh TAP growth medium (control condition) for 6 hrs. 200 $\mu$l of  cultured cells for each experimental condition were seeded in a 96 well and time-lapse imaged immediately. For study of motility at different densities,  200 $\mu$l of  cultured cells ($2 \cdot 10^6$ cells/ml and $10^6$ cells/ml respectively) were let sediment on the bottom of a 96 well before imaging. Images were acquired with DMi8 (Leica) using bright field objective at 20x at $0.5-6$ frames per second.

\subsubsection{Image segmentation. }
Image segmentation is done using standard Matlab functions for the image processing.
First, the image is thresholded using “edge” function in two steps: the automatic threshold is identified and then lowered to achieve better edge detection. The detected edges are dilated using “imdilate“ function. Then, the closed areas are filled with imfill function to remove black spots inside detected cells. Finally, bwareaopen is used to remove small noise. The result is a 
mask that selects only regions occupied by algae.

\subsubsection{Particle image velocimetry (PIV). }
The measurements of the velocity field were done using the PIVlab app for Matlab \cite{thielicke2014pivlab}. The method is based on the comparison of the intensity fields of two consequent photographs of algae. The difference in the intensity is converted into velocity field measured in $px/frame$ and then converted to $\mu m/h$ \cite{chepizhko2016bursts}. To avoid
spurious noise PIV was applied after image segmentation only to the regions occupied by algae.

\subsection{Simulations}
\subsubsection{Model for interacting 2D active disks. }
We performed Molecular Dynamics simulations of a system made of active self-rotating particles using LAMMPS (Large-scale Atomic/Molecular Massively Parallel Simulator) \cite{plimpton1993fast}.
Each particle (described as a 2D disk) has the following physical properties:
\begin{itemize}
\item it is an active particle and it has the ability of both self propel to move in a straight direction and to self-rotate around its center
\item it is moving on a viscous medium, so performing a Brownian motion
\item the system is made up of particles with two different radii in order to avoid crystallization
\item it interacts with its nearest neighbor particles via a granular potential (Hertzian potential) to which we add a cohesion term (Derjagun-Muller-Toporov Model \cite{derjaguin1975effect}). This is done in order to take into account the adhesion properties of cells. 
\end{itemize}
The equations of motion describing the dynamics of the set of two-dimensional disks at time $t$ are given by
\begin{equation}
\frac{d^2 \overrightarrow{x}_i (t)}{dt}=\frac{1}{m} [\overrightarrow{\Gamma}_i(t)+\overrightarrow{\chi}_i(t)+\overrightarrow{\Phi}_i(t)+\sum_{n.n.}\overrightarrow{\Psi}^n_{ij}(t)+\sum_{n.n.}\overrightarrow{\Lambda}_{ij}(t)]
\label{Eq1}
\end{equation}
\begin{equation}
\frac{d^2 \theta_i (t)}{dt}=\frac{1}{I_z} [\overrightarrow{\Upsilon}_i(t)+\overrightarrow{\Theta}_i(t)+\overrightarrow{\tau}_i(t)+\sum_{n.n.}-( \overrightarrow{R}_i \times \overrightarrow{\Psi}^t_{ij}(t))]
\label{Eq2}
\end{equation}
The quantities are dimensionless because units are set to Lennard-Jones units in LAMMPS. It is
however always possible to convert to real units by setting the disk radius to
the typical algae radius.
Since particles are moving in a viscous environment, the term $\overrightarrow{\Gamma}_i(t)$ takes into account the friction due to the surrounding medium. Its value is:
\begin{equation}
\overrightarrow{\Gamma}_i(t)=-\frac{m}{\beta}\overrightarrow{v} (t)
\end{equation}
As we can see this term is proportional to the linear velocity of the particle and to the particle mass $m$. Here, the input parameter $\beta$ is inversely proportional to the fluid viscosity. Since cells moving in a medium are usually in an overdamped regime,  we set $\beta=0.1$ so that Brownian dynamics can effectively be considered as an overdamped Langevin dynamics.
Friction enters also in the rotational motion of our $2D$ disks, since we have a friction term proportional to the angular velocity and to the moment of inertia of the disks $I$ given by
\begin{equation}
\overrightarrow{\Upsilon}_i(t)=-\frac{10}{3} \frac{I}{\beta}\overrightarrow{\omega} (t)
\end{equation}
Next, we consider in Eqs. \ref{Eq1} and \ref{Eq2} a term representing random noise, described by $\overrightarrow{\chi}_i(t)$ and $\overrightarrow{\Theta}_i(t)$. From the fluctuation/dissipation theorem, the magnitude of $\overrightarrow{\chi}_i(t)$  is proportional to $\sqrt{\frac{K_{b} T m}{dt \beta}}$, where $K_b$ is the Boltzmann constant, T is the desired temperature, $m$ is the mass of the particle, $dt$ is the timestep, and $\beta$ is the damping factor. In the case of $\overrightarrow{\Theta}_i(t)$, the mass in the previous equation is substituted by the moment of inertia $I$.  The random noise is uncorrelated with $\langle \xi(t) \rangle=0$ and $\langle \xi(t) \xi(t') \rangle \propto \delta (t-t')$.

Since our aim is to describe active particles, we include in Eq. \ref{Eq1} and Eq. \ref{Eq2}  terms taking into account the ability of the particle to self sustain its motion.
The term $\overrightarrow{\Phi}_i(t)$ takes into account the ability of the cell to self-propel. The biological mechanisms that allow the cell to move are simply modeled as a force with a constant modulus  pointing in the direction in which the particle was already moving,
considering a characteristic time until the particle can change direction 
\begin{equation}
\overrightarrow{\Phi}_i(t)=f_{0}\frac{\overrightarrow{v} (t)}{\vert v \vert}
\end{equation}
The term $\overrightarrow{\tau}_i(t)$ takes into account  the self-rotation of the particle
adding a constant torque along the $\widehat{z}$ direction at each time step for each particle. Since at the beginning of the simulation, particles are endowed with an initial angular velocity, either clockwise or counter-clockwise, the torque term initially follows the direction of rotation, so that a particles that moves clockwise at the beginning continues to rotate in that direction under the self rotation effect. Of course changes of rotation direction can arise due to interaction among particles.
In order to study the effect of rotation on the collective properties of our system, we performed simulations for different values of the active torque $\tau$.
The term $\overrightarrow{\Psi}_{ij}(t)$ describes contact interactions between the particles, according to the Hertzian model \cite{brilliantov1996model,silbert2001granular}. In particular the form of the force is:
\begin{equation}
\begin{split}
\overrightarrow{\Psi}_{ij}(t)=&\sqrt{\delta}\sqrt{\frac{R_{i}R_{j}}{R_{i}+R_{j}}}[(k_{n}\delta\overrightarrow{n}_{ij}-m_{eff}\gamma_{n}\overrightarrow{v}_{n})+\\
&-(k_{t}\overrightarrow{\Delta s_{t}}+m_{eff}\gamma_{t} \overrightarrow{v}_{t})]
\end{split}
\end{equation}
Where $R_{i}$ and $R_{j}$ are the radii of the interacting particles. In our simulations,  particles have two different diameters $R_{1}=1.96$ and $R_{2}=1.4$, in order to avoid crystallization. The force is divided in two components, the force that is normal to the contact surface between the two particles and the one that is tangential.
The normal force has two terms, a contact force and a damping force. Here $\delta$ is the overlap distance between two particles, $k_{n}$ is the elastic constant for the normal contact, $\overrightarrow{n}_{ij}$ is the unit vector along the line connecting the centers of the two interacting particles, $\gamma_{n}$ is the viscoelastic damping constant for normal contact and $\overrightarrow{v}_{n}$ is the normal component of the relative velocity of the two particles.
This component of the force enters in \ref{Eq1}, influencing the equation of motion of the center of mass of the disk.  The tangential force also has two terms: a shear force and a damping force. The shear force contains a "history" effect that accounts for the tangential displacement between the particles for the duration of the time they are in contact. 
Here $k_{t}$ is the elastic constant for tangential contact and $\overrightarrow{\Delta s_{t}}$ is the tangential displacement vector between two particles. In the tangential damping force the term  $\gamma_{t}$ is the viscoelastic damping constant for tangential contact and $\overrightarrow{v}_{t}$ is the tangential component of the relative velocity of the two particles. 
This tangential component of the Hertzian interaction enters in the rotational 
equation of motion, via the torque $-\overrightarrow{R}_i \times \overrightarrow{\Psi}^t_{ij}(t)$.

Then, in order to study the role of cohesion between particles, typical of many biological systems, a term taking into account the adhesion is inserted in Eq. \ref{Eq1}.
This is done using the Derjaguin-Muller-Toporov model \cite{derjaguin1975effect,barthel2008adhesive}, where the adhesive force has the form $\overrightarrow{\Lambda}_{ij}(t)=-\frac{A_{cc} R_{ij}}{6 \epsilon^{2}}$, where $A_{cc}$ is the Hamaker constant that takes into account the coefficient of the particle-particle pair interaction, $R_{ij}=\frac{R_{i}+R_{j}}{R_{i}R_{j}}$ is the effective radius of the two touching particles and $\epsilon$ is the least possible spacing between the contact surfaces.
In particular the adhesion force is calculated in the LAMMPS code \textit{pair-dmt} distinguishing between two cases:
If the distance among the centers of two spheres is bigger than the sum of the radii $r>a_{1}+a_{2}$, the adhesion force takes the form  $\overrightarrow{\Lambda}_{ij}(t)=-\frac{A_{cc} R_{ij}}{6 [r-(a_{1}+a_{2})+\epsilon]^2}$, otherwise
 the adhesion force becomes $\overrightarrow{\Lambda}_{ij}(t)=-\frac{A_{cc}R_{ij}}{6 \epsilon^2}$. So the adhesion is simply represented by a spring force when two particles overlap. In the following we studied the role of adhesion performing simulations with different values of the Hamaker constant $A_{cc}$. Integrating the Equations \ref{Eq1} and \ref{Eq2} it is possible to update positions and velocities of the particles at each time step of our simulation.

\subsubsection{Parameters used for the simulations. }
For the simulations we used 2D disks of radius $R_1=1.96$ and $R_2=1.4$ and density of the disk $d=0.46$ (so mass $m=d*\pi R^2$). We used a time step of $0.0001$ in LJ units and run the simulations for $10^6$ steps, so covering $100$ time units. The temperature is constant during the integration of the equation of motion and equal to $T=1$.

\section{Results}

\subsection{PIV analysis of \textit{C. reinhardtii}}
\begin{figure*}[htb]\centering 
\includegraphics[width=16cm]{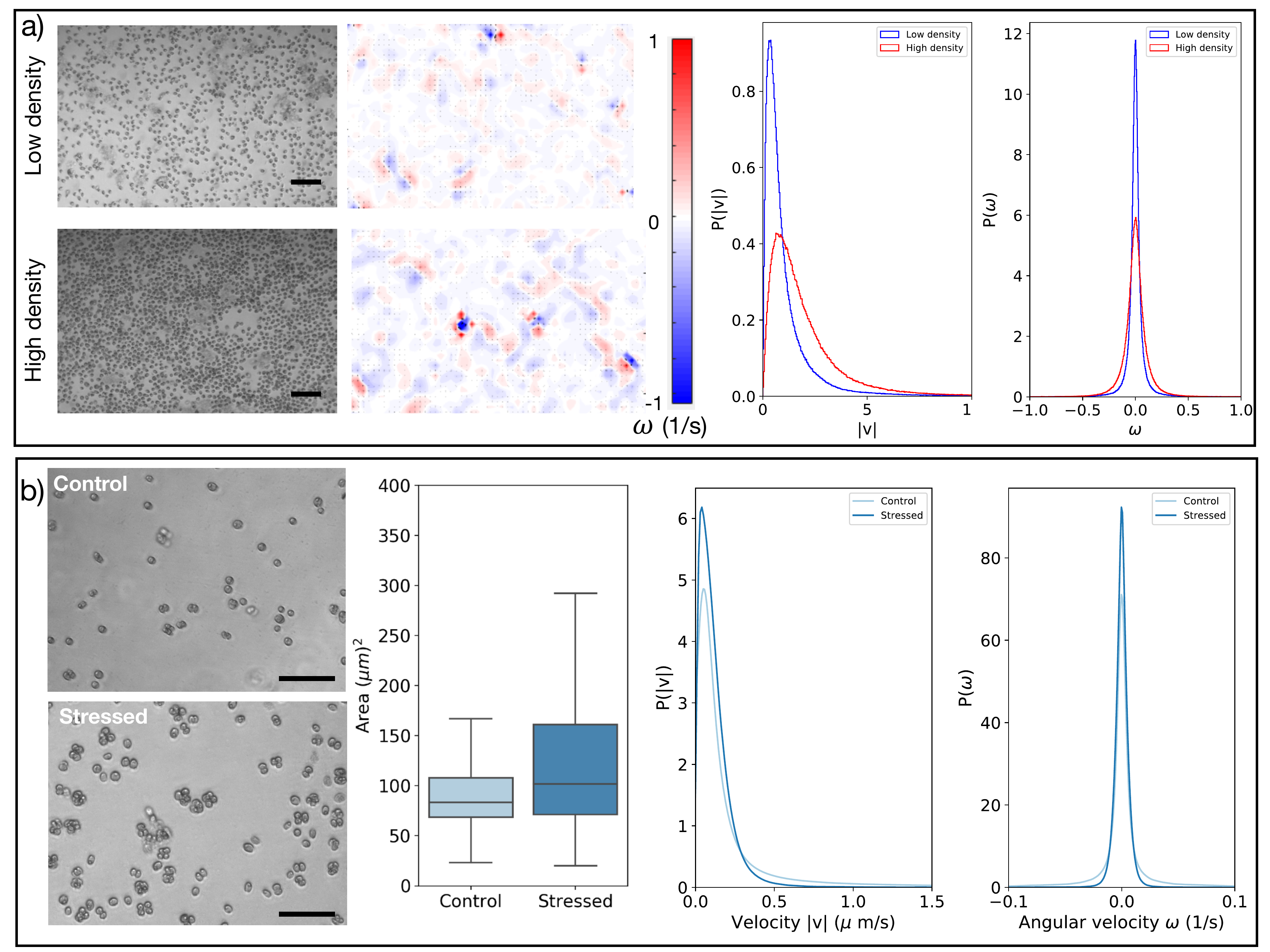}
 \caption{\label{fig:1} a) Typical snapshots of \textit{C. reinhardtii} suspensions at different densities (scale bars $100\mu$m) are shown together with the corresponding maps of angular velocities obtained by PIV. On the right-hand side, we report plots of the probability distribution of the angular velocity $P(\omega)$ and of the absolute value of the linear velocity $P(|v|)$.  b) Images of the algae (scale bars $50\mu$m) at low density with and without the stress induced by the presence of NaCl are shown together with a box plot of cluster areas,
the distribution of angular velocities $P(\omega)$ and of the absolute value of the linear velocities $P(|v|)$. }
 \end{figure*}

The PIV analysis performed on time-lapse videos of \textit{C. reinhardtii} allows us to reconstruct the distribution probabilities for linear and angular velocities (Fig. \ref{fig:1}a). The plots in 
Fig. \ref{fig:1}a show the distributions of the absolute value of the linear velocities, following a Rayleigh-like distribution, while the angular velocity distribution has a Gaussian-like behavior. We notice that increasing the density, the algae became more motile, since the linear velocity peak is shifted toward higher values and the variance of the angular velocity distribution increases. Another important detail is that the fraction of regions with zero angular velocity decreases when increasing the density. This is also evident from the angular velocity maps obtained by the PIV analysis shown again in Fig. \ref{fig:1}a, where it can be observed that the algae rotate and tend to form rotating clusters and vortex like behavior. The formation of those rotating clusters is clear also
observing algae at very low density stressed with $150mM$ NaCl added to the medium
(Fig. \ref{fig:1}b). The stress agent enhances this behavior since algae tend to aggregate in response to it. The probability distribution obtained in the more stressed phase indicates 
that aggregation leads to a decrease of velocity and angular velocity fluctuations.

\subsection{Simulations at low density }
\begin{figure*}[htb]\centering 
\includegraphics[width=16cm]{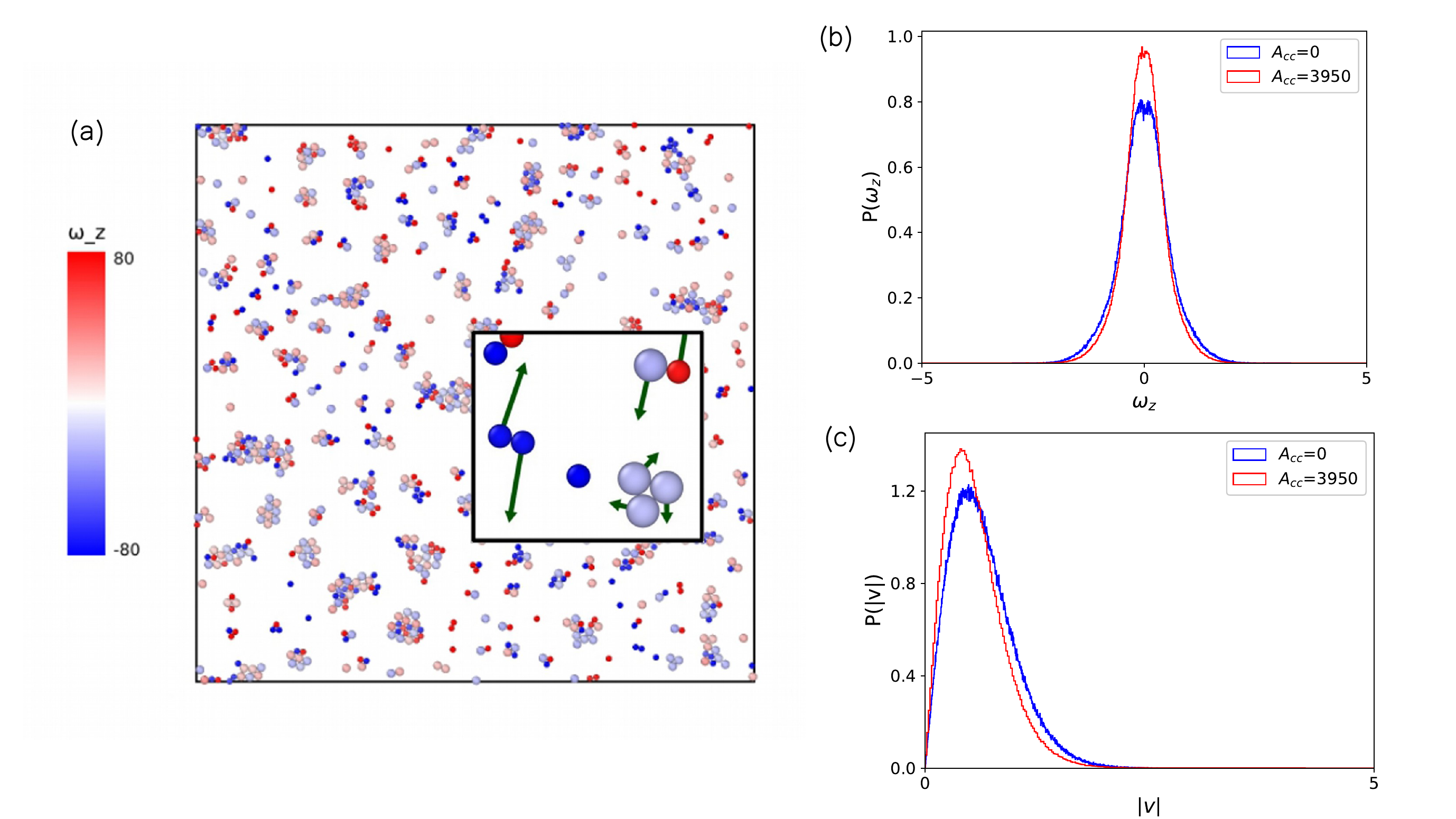}
 \caption{\label{fig:5} Results of the simulations at low density $\phi=0.14$ in presence of active torque are illustrated. In $5$a the system of $1000$ particles simulated with LAMMPS is shown, with an active torque of $\tau_z=6000$. Here the cohesive potential is switched on (with the following parameters: $A_{cc}=3950$ and cutoff$=\frac{3}{2}(R_1+R_2)$). The particles are colored according to their angular velocity $\omega_z$. The formation of clusters is clearly visible. In the inset a zoom of a portion of the simulated box is shown, together with the velocity vectors of the particles. In particular it is easy to see a system of three collectively rotating particles. In $5$b the probability distribution of the angular velocity is shown for the system in presence of the adhesion force ($A_{cc}=3950$) and without cohesion among disks respectively. In $5$c the probability distribution of the linear velocity is shown as before in presence of the adhesion force ($A_{cc}=3950$) and without cohesion among disks. In both $5$b and $5$c the active torque is $\tau_z=10$. }
 \end{figure*}

In order mimic what is observed in \textit{C. reinhardtii} and to gain insight on those systems of active rotators, we performed simulations in LAMMPS using the model described in the model section. We fist consider the low density case and in Fig. \ref{fig:5}a we can see an example of a system with packing fraction $\phi=\frac{V_{particles}}{V_{TOT}}=0.14$, in which we have both active torque $\tau_z=6000$ and cohesion with $A_{cc}=3950$. Particle start at random positions and after a while they aggregate into clusters with a size determined by the cutoff of the adhesion potential. The interesting property of those cluster is that they rotate collectively. This behavior is similar to what observed in \textit{C. reinhardtii} and also in previous studies on 2D active spinners embedded in passive colloidal monolayers \cite{aragones2019aggregation}. In the latter case, it has been observed that the presence of a passive monolayer that behaves elastically as a solid-like material, induces an attractive interaction between the active rotating particles, which results in aggregation of spinners \cite{aragones2019aggregation}. In our case, what we noticed is that switching off the adhesion among the particles, the clusters do not form any more.

From the simulations, we also extracted the probability distributions for the linear and angular velocities of the particles, in the cases with and without the adhesion force, shown in 
Fig. \ref{fig:5}b and Fig. \ref{fig:5}c. The distributions we obtained from the model are very similar to those obtained from the experiments on natural active rotators such as \textit{C. reinhardtii}, suggesting that our model can capture some important features of those biological systems. We observed that the distribution of the angular velocity shows a Gaussian-like shape,  with and without cohesion. The distribution of the absolute value of linear velocities is instead, well described by a Rayleigh-like distribution for the case without cohesion, while the fit is poor in the cohesion case.

\subsection{Simulations at the jamming-unjamming transition}
After comparing model and experiments in the dilute regime, we investigate how the self-rotation of the disks can affect the phase behavior of a dense system of interacting rotators. To improve the statistics, our results are averaged over ten initial configurations
initially placed into the jammed phase. To do this we created a simulation box with 1000 disks in random positions (500 with radius $R_1=1.96$ and 500 with radius $R_2=1.4$, 
corresponding to binary mixture with diameter ratio of $1.4$ as already done in previous studies on jamming and packing of 2D disks \cite{o2002random,o2003jamming}). We then performed subsequent steps of box reduction and energy minimization, monitoring the behavior of the pressure of the system as a function of the packing fraction $\phi$. 

It is already known from previous studies on random packing of frictionless particles that $\phi$ at which the pressure becomes non zero is the same as the jamming threshold, when also the static shear modulus becomes non-zero \cite{o2002random}. This can be understood by thinking that when the packing fraction is small, particles do not touch and the internal pressure is zero.  Increasing the particle density via box reduction, the system reaches a state in which the particles touch and are blocked into a rigid structure. At this point, a further decrease
of the packing fraction will lead to a pressure increase. Hence, pressure is a good indicator of the jamming point. In our case, we observe that the pressure is zero until $\phi \simeq 0.78$ and increasing rapidly for larger values of $\phi$. To ensure that the system is in the jammed
state, we chose initial configurations with $\phi=0.87$. Previous work on packing and jamming of 2D bidisperse hard disks at $T=0$ shows that  the value for random close packing $\phi_{RCP}$ (maximum density without crystallization) is $\phi=0.84$ \cite{o2003jamming}. 
In our case, several factors have to be taken into account. First,  we are considering friction where usually $\phi_J < \phi_{RCP}$, so the jamming is reached before the random close packing density. Furthermore, we are considering a system that is not at zero temperature. Thus as suggested by the jamming diagram proposed by Liu e Nagel \cite{liu1998jamming}, the jamming transition is expected to occur at a higher density. These facts justify the value we found for the onset of jamming.  

It has recently been observed that the critical value of the density at which the jamming transition takes place can depend also on the conditions in which the system has been prepared and varies also within the same material \cite{kumar2016memory} making it hard to find a
well defined value of the packing fraction for the onset of jamming.
Furthermore, we have to consider that our 2D disks are not hard, their 'hardness' being controlled by the $k_n$ coefficient in the Hertzian potential. Thus, they can overlap and elastically deform, reaching higher values of packing fractions.

\subsubsection{The role of self-rotation. }
\begin{figure*}[htb]\centering 
\includegraphics[width=18cm]{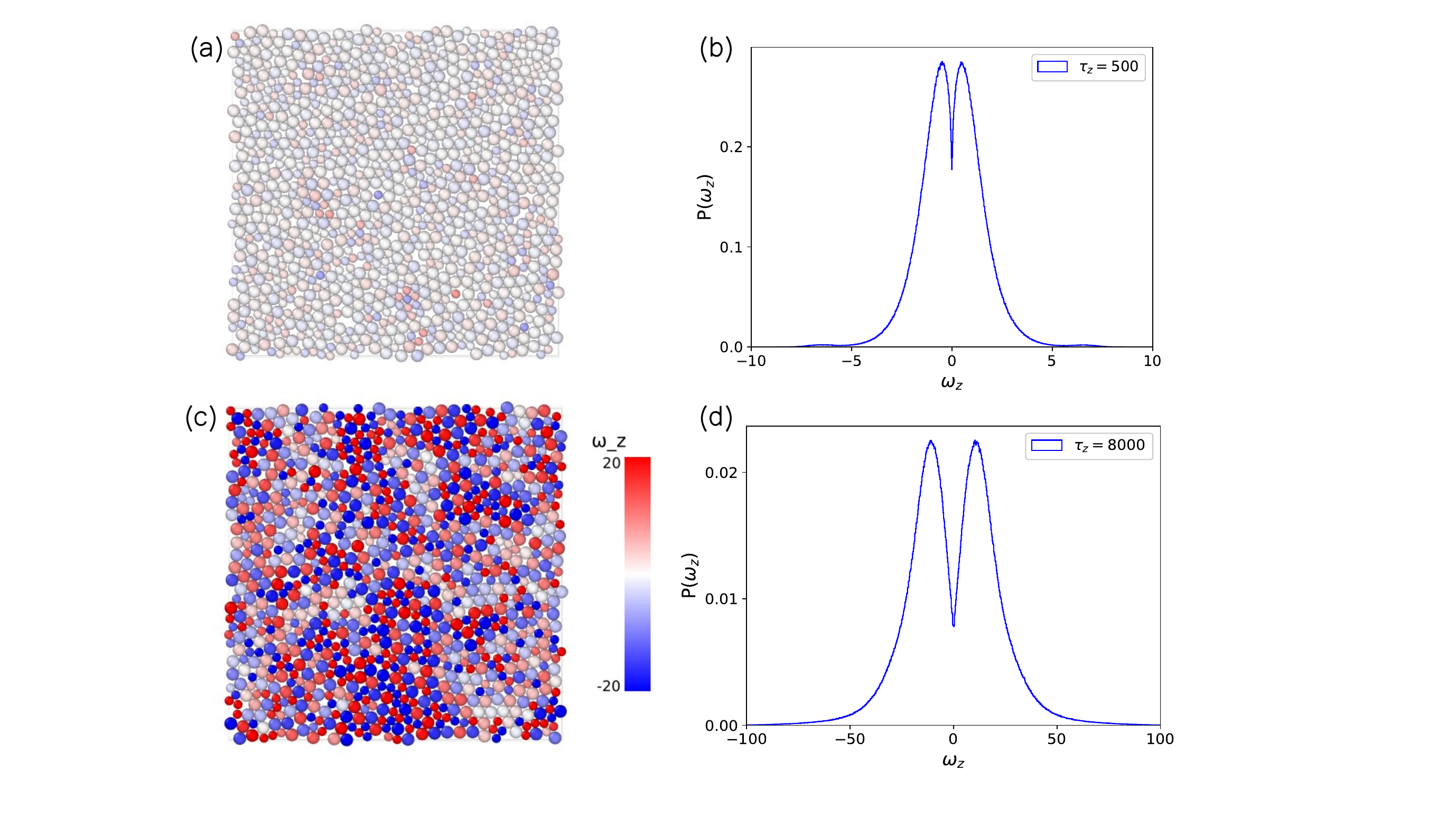}
 \caption{\label{fig:2} a) A typical snapshot of the system in the jammed phase, with $\tau_z=500$, while in b) the system is in the unjammed phase with $\tau_z=8000$. The 
 disks are colored according to their angular velocity $\omega_z$; the same color coding has been used for the two images. c) The probability distribution of the $z$ component of the angular velocity for the system in the jammed phase $(\tau_z=500)$. The probability is averaged on time and on the $10$ different initial configurations. d) The probability distribution of the $z$ component of the angular velocity for the system in the unjammed phase $(\tau_z=8000)$. The probability is averaged on time and on the $10$ different initial configurations. In this case, 
 it is clear the presence of two peaks due to the active torques injected in the system $\tau_z=\pm8000$.}
 \end{figure*}

\begin{figure*}[htb]\centering 
\includegraphics[width=16cm]{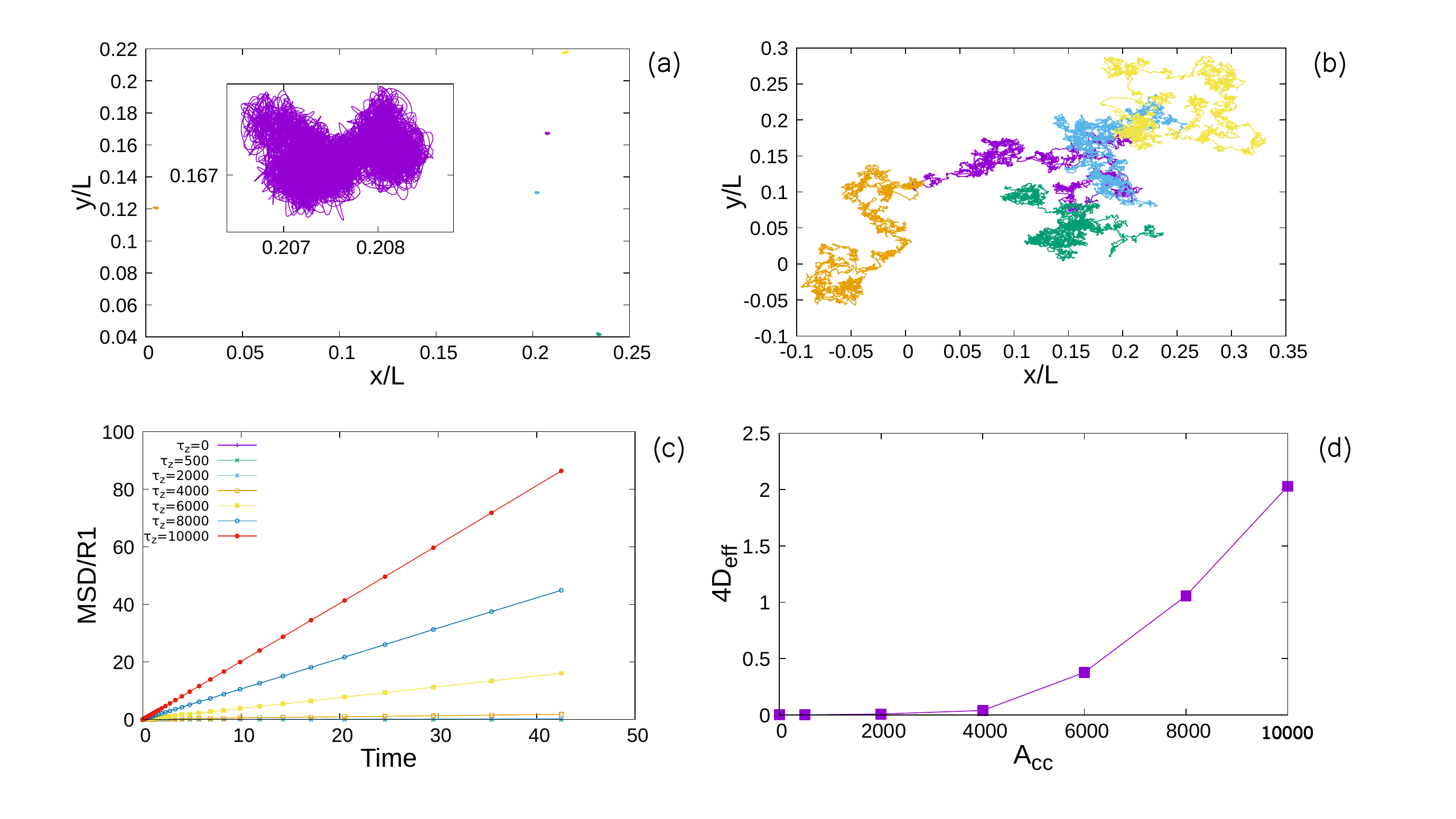}
 \caption{\label{fig:3} a) Trajectories of five randomly selected disks from the system in the jammed phase $(\tau_z=500)$. The displacement is very small so that a zoom on a single disk trajectory is shown in the inset. b) Trajectories of five disks randomly selected from the system in the unjammed phase $(\tau_z=8000)$. In both plots, the the coordinates are rescaled using the length of the simulation box $L$. c) The time evolution of the mean-square displacement averaged over all the disks belonging to the system is shown for different values of the active torque, showing a clear increase for increasing values of the disks self rotation.
d) The diffusion coefficient, obtained from a linear fit of the long time region of the mean square displacement is plotted. The increase after a critical value of the active torque $\tau_z$ is associated to a phase transition from a jammed/solid-like phase to an unjammed/flowing phase of the system.}
 \end{figure*}

We perform MD simulations starting from the initial configurations and changing the value of the active torque $\tau_z$. When self-propulsion is switched off ($f_0=0$) as well as the adhesion term ($A_{cc}=0$), we noticed that at low active torques the system stays in a jammed phase, characterized by a very low mobility of the active disks. Above a critical value of the active torque, the system switches to a flowing, unjammed phase, characterized by a high mobility of the disks. We analyzed the probability distribution of angular velocities and in Fig. \ref{fig:2}  we observe as the single particles increase their angular velocity due to the application of the active torque, as expected. The two peaks clearly visible in the probability distribution reflect the torque injected in the system, equal to $\pm \tau_z$. But to clearly gain insight on what is happening at the phase transition, we studied individual trajectories of the disks,
computing their mean square displacement (MSD). In Fig. \ref{fig:3}, we observe that in the jammed phase trajectories are localized while for high values of the active torque they diffuse. The mean square displacement stays close zero for low torques while it grows linearly for larger values of the self-rotation. From a linear fit of the long time region of the MSD, we also extracted an effective diffusion coefficient, that clearly show a sharp increase at ($\tau_z=4000$), suggestive of a phase transition into a flowing state.

\subsubsection{The role of adhesion. }
Since we have seen that the self-rotation can lead from a jammed to an unjammed state, it is interesting to investigate the role of adhesion, present in many cellular systems, including
\textit{C. reinhardtii} where it is triggered by stress. As discussed in the Model section,  adhesion is modeled using the Derjaguin-Muller-Toporov model and the parameter used to tune the intensity of the adhesion force in the simulations is the Hamaker constant $A_{cc}$. We performed simulation with $\tau_z=6000$, so in the unjammed phase switching on the adhesion term. Analyzing the mean square displacement and as before the effective diffusion coefficient, it emerges that increasing the adhesion strength the system remains unjammed until a critical value ($A_{cc}\sim1500-2000$) at which diffusion is strongly reduced, unveiling a transition to a jammed phase.

\begin{figure*}[htb]\centering 
\includegraphics[width=16cm]{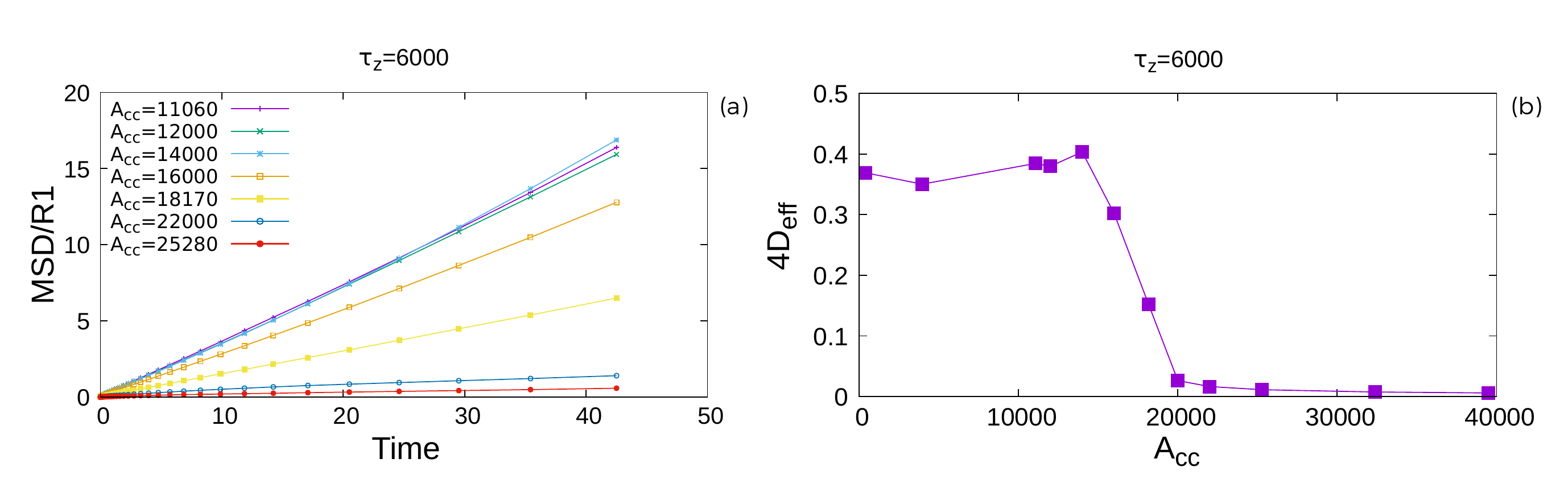}
 \caption{\label{fig:4} a) The mean square displacement of the system in the unjammed phase $(\tau_z=6000)$ and high density ($\phi=0.87$) is shown for different values of the Hamaker constant $A_{cc}$, that defines the intensity of the cohesive potential among the particles.  Increasing the strength of the cohesion, the mean square displacement decreases. b) The effective diffusion coefficient averaged on all the disks is extracted from the linear fit of the mean square displacement. Here the phase transition from the unjammed to the the jammed phase is clearly visible as the diffusion coefficient rapidly drops to zero for large adhesion strengths. }
 \end{figure*}

\section{Discussion}
Our work was inspired by the observation of a natural example of active rotators like 
\textit{C. reinhardtii}. Quantification by image segmentation and PIV analysis shows
that this kind of algae can not only self-rotate  \cite{choudhary2019reentrant} but also aggregate forming collectively rotating clusters. The formation of these aggregates is observed both in high density limit, and low density in presence of a stress agent, such as NaCl.
Starting from these simple observations, we built and simulate a model of 2D active disks that have the ability to self-rotate and interact with each other. We found that in the low density limit ($\phi=0.14$) and in presence of an adhesion term among the disks, the active rotators tend to aggregate and form rotating clusters, in analogy with what is observed in algae. In particular, we saw that the adhesion term plays a crucial role in the formation of clusters, suggesting that a form of attraction should also be present in the case of \textit{C. reinhardtii}.

Furthermore, we studied with our simulations the role of self-rotation in a jammed system. We observed that self-rotation alone can lead to a phase transition from a jammed solid-like state to an unjammed, flowing phase. In the past the role of active forces has been investigated, studying  the phase diagram of 2D soft disks, that exhibit a liquid phase with giant number fluctuations at low packing fraction  $\phi$ and high self-propulsion speed $v_0$ and a jammed phase at high $\phi$ and low $v_0$ \cite{henkes2011active} or in active dumbbell systems with different packing fraction and P{\'e}clet number \cite{cugliandolo2017phase}. These studies, however, did not
consider self-rotations as we did. We also studied the effect of adhesion 
and investigated its role when combined to the self-rotation of the diks.
We revealed how adhesion can act in the opposite direction with respect to self-rotation, 
promoting jamming. Increasing the adhesion strength, we can move the system from a flowing unjammed phase to a jammed one.  It would be interesting to observe a similar phase transition 
in experiments controlling self-rotation and adhesion, for instance in chiral ative fluids made of superparamagnetic particles in a magnetic field \cite{aragones2019aggregation}).

A better characterization of the key features of this jamming-unjamming transition in active particles systems could be very useful to better understand biological processes in which the cells involved have the ability of self-rotate, both at a single particle level or as collective rotation. For example, experiments involving epithelial cells confined in narrow channels
showed the formation of vorticity, suggesting a possible role for rotations in collective
cell migration \cite{vedula2012emerging}.  Another context in which the mechanical properties of tissues gain a peculiar interest is in the study of cancer cells, and in particular in the formation of metastasis. It has been observed that cancer cells are softer than non-cancerous ones \cite{oswald2017jamming}, divide more often than healthy cells and, as in the case of the epithelial-to mesenchymal transition (EMT), they decrease the cell-cell adhesion, potentially
allowing for rotational motion. All those features contribute to fluidize a confluent tissue of cancer cells, favoring the unjamming transition and so the formation of diffusing groups of cells (for a review see \cite{laporta2019}). Hence, our theoretical study of a model system of active rotators reveals how self-rotation of the active particles is a parameter that can control the jamming-unjamming transition, besides already well studied mechanisms such as
self-propulsion \cite{henkes2011active} or density, and can help in better understanding
physical aspects of cancer cell invasion \cite{laporta2017}.

\section*{Acknowledgements}
S. Z. thanks the Alexander von Humboldt foundation for the Humboldt Research Award for 
support and Ludwig-Maximilian University and 
Friedrich-Alexander-Universit\"at Erlangen-N\"urnberg for hospitality.


\providecommand*{\mcitethebibliography}{\thebibliography}
\csname @ifundefined\endcsname{endmcitethebibliography}
{\let\endmcitethebibliography\endthebibliography}{}

\end{document}